\documentclass[aps,superscriptaddress,twocolumn,floats,prd,showpacs,showkeys]{revtex4} 
\usepackage{graphicx} %
\usepackage{epsfig,amsmath}
\usepackage{amssymb}
\usepackage{rotate}
\usepackage{color}
\usepackage[ps2pdf=true]{hyperref}
\usepackage{bm}

\DeclareFontFamily{OT1}{pzc}{}
\DeclareFontShape{OT1}{pzc}{m}{it}%
            {<-> s * [1.10] pzcmi7t}{}
\DeclareMathAlphabet{\mathscr}{OT1}{pzc}%
                                {m}{it}

\newcommand{\be}{\begin{equation}}
\newcommand{\ee}{\end{equation}}
\newcommand{\bea}{\begin{eqnarray}}
\newcommand{\eea}{\end{eqnarray}}
\def\ba#1\ea{\begin{align}#1\end{align}}

\newcommand{\reffig}[1]{Fig.~\ref{fig:#1}}

\newcommand{\vlos}{v_{\rm los}}

\newcommand{\D}{\Delta}

\newcommand{\Mpch}{\,{\rm Mpc}/h}

\newcommand{\Msunh}{\,M_{\odot}/h}

\newcommand{\s}{\sigma}

\begin{document}

\title{Testing Gravity with the Stacked Phase Space around Galaxy Clusters}

\author{Tsz Yan Lam}
\affiliation{Kavli Institute for the Physics and Mathematics of the Universe
(Kavli IPMU), University of Tokyo, Chiba 277-8583, Japan}

\author{Takahiro Nishimichi}
\affiliation{Kavli Institute for the Physics and Mathematics of the Universe
(Kavli IPMU), University of Tokyo, Chiba 277-8583, Japan}

\author{Fabian Schmidt}
\affiliation{Theoretical Astrophysics, California Institute of
    Technology, Mail Code 350-17, Pasadena, California  91125}

\author{Masahiro Takada}
\affiliation{Kavli Institute for the Physics and Mathematics of the Universe
(Kavli IPMU), University of Tokyo, Chiba 277-8583, Japan}

\begin{abstract}
In General Relativity, the average velocity field of dark matter
around galaxy clusters is uniquely determined by the mass profile. The
latter can be measured through weak lensing. We propose a new method
of measuring the velocity field (phase space density) by stacking
redshifts of surrounding galaxies from a spectroscopic sample. In
combination with lensing, this yields a direct test of gravity on
scales of 1-30 Mpc. Using N-body simulations, we show that this method
can improve upon current constraints on $f(R)$ and DGP model
parameters by several orders of magnitude when applied to upcoming imaging and redshift surveys.
\end{abstract}

\maketitle


The accelerated expansion of the Universe is the most tantalizing
problem in modern cosmology. Within Einstein's
General Relativity (GR), 
the cosmic acceleration can be explained by
introducing a mysterious smooth component, dark energy.
However, it can also be interpreted as signature of the breakdown of GR on
cosmological scales.  Many on-going and upcoming wide-area galaxy
surveys aim at testing dark energy and modified gravity scenarios as
the origin of cosmic acceleration.

Cosmological probes of gravity are based on reconstructing
the perturbations in the space-time metric and their
relation to matter \cite{Bertschinger,JainZhang}.  
Weak gravitational lensing provides a clean measurement of the
lensing potential, while the timelike potential can be probed
through the modulations in redshift
caused by peculiar velocities of galaxies.  
In this {\em Letter}, we propose a new method of testing gravity at
intermediate scales ($1 - 30$~Mpc) by measuring a projection
of the position and velocity space (hereafter phasespace)
around massive galaxy clusters.  
If GR is valid, the phasespace around the sampled clusters is uniquely determined 
by the mass density profile, which can be measured through stacked
weak lensing.  
In other words, comparing the measured mass density and velocity
profiles allows for a model-independent test of Einstein gravity.  
The scales probed are
complementary to and potentially provide more information than 
the linear regime studied in most previous studies \cite{ZhangEtal,ReyesEtal}, 
or the small scales considered in 
\cite{SchwabEtal09,Schmidt10,Wojtaketal:11,boyarskyruchayskiy11}.  
Moreover, this test is a \emph{generic} probe of
gravity:  adding other, non-standard ingredients such as massive
neutrinos or primordial non-Gaussianity, for example, will likely have a 
negligible impact on the relation between mass density and velocity
profiles.  This is not the case for other commonly considered probes of 
gravity, such as the matter power spectrum or cluster abundance.  
The main challenge lies in modeling the observables on these scales.  
We will demonstrate the feasibility of our method by using N-body simulations
for Einstein and modified gravity models.

\label{sec:ps}

{\em Methodology}: Consider a sample of galaxy clusters (with accurate
redshifts) in a
cosmological volume covered by a spectroscopic galaxy survey.  We can
then construct the two-dimensional distribution of galaxy-cluster pairs in
terms of the transverse distance $r_p$, and the relative line-of-sight
velocity $\vlos$.  More precisely, we have 
\ba r_p =
d_A(z_c)\Delta\theta_{gc},\hspace{1em}
\vlos =
c (z_g-z_c), 
\ea 
where
$\Delta\theta_{gc}$ is the angular separation of galaxy and cluster,
$d_A$ is the comoving angular diameter distance, $c$ is the speed of
light, and $z_g,\:z_c$ denote the galaxy and cluster redshifts,
respectively. The average phase space distribution is estimated by stacking all 
cluster-galaxy pairs.

\begin{figure}[t!]
\centering
\includegraphics[angle=-90,width=0.47\textwidth]
{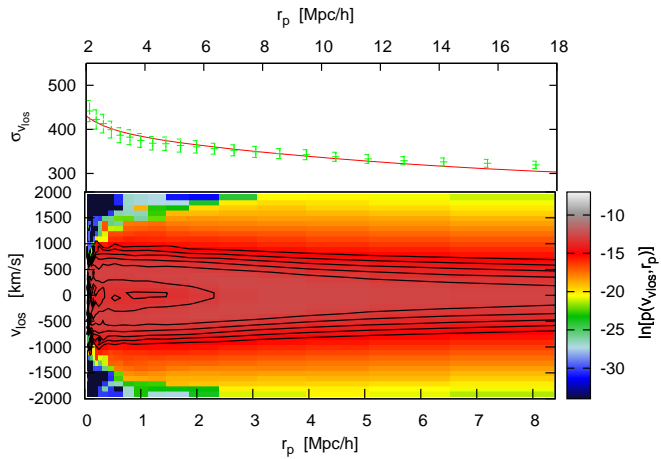}
\caption{{\em Lower panel}: 
The $\vlos-r_p$ phase space distribution (in logarithmic scale) as measured
using halo catalogs constructed from N-body simulations in $\Lambda$CDM
(see also \cite{Wojtaketal:11}); 
we considered primary halos (``clusters'') with masses $\ge 10^{14}M_{\odot}/h$ 
and secondary halos (``galaxies'') in the
range $3\times 10^{13}\le M\le 10^{14}~M_\odot/h$. 
{\em Upper panel}: The dispersion of the line-of-sight velocity
 distribution $\sigma_{\vlos}$ as function of $r_p$. 
The data points with error bars
 are computed from the simulation results in the lower panel, while the
 solid curve is our analytical model prediction. The error bars are
scaled 
to mimic
 the measurement accuracies for a spectroscopic 
survey of 2000 sq. degrees over
 $0.2<z<0.4$. 
\label{fig:ps}}
\end{figure}

The lower panel of \reffig{ps} shows
this distribution, using only peculiar motions, measured in the
Einstein-gravity N-body simulations of \cite{NishimichiTaruya:11} 
around halos with masses $M\ge
10^{14}\Msunh$ identified at $z=0.35$, where we assumed a concordance $\Lambda$ and cold dark
matter cosmological model ($\Lambda$CDM).  We use the output at $z=0.35$
of 20 simulations of $1.5~({\rm Gpc}/h)^3$ volume each.  
We defined halos using the
friends-of-friends finder algorithm with linking length 0.2 times the
mean particle separation, and assigned center-of-mass positions and
velocities using the member particles.  
To mimic a galaxy redshift survey, we select secondary halos in the
mass range $3\times 10^{13}\le M_{s}<10^{14}M_\odot/h$ as galaxies in
a cube of side length 40~Mpc$/h$ centered on each primary halo.  
In real galaxy surveys, such a selection in real space is not possible
of course; we will return to this point below.  
By stacking over many clusters and binning in $r_p$, we average
over triaxial or irregular density profiles, yielding a distribution 
which is only a function of $r_p$ and $|v_{\rm los}|$.    

The lower panel of 
Fig.~\ref{fig:ps} clearly shows two distinct regimes; at small radii
$r_p\lesssim 2\Mpch$, iso-density contours are closed,
while on larger scales the contours become open, reflecting the ongoing
infall onto the massive halos.  The boundary between these two regimes
has been used in the caustic method \cite{Diaferio}.  
As shown in a forthcoming paper, we can construct an accurate model
of the $\vlos-r_p$ distribution of dark matter halos through
a combination of N-body simulations and analytical theory.  
The upper panel shows the analytical prediction for the RMS dispersion
of $\vlos$ as a function of $r_p$ (estimated through the standard sample RMS). 
The model prediction is in good agreement with the
simulation result, within the statistical errors of the simulation
measurements, which were measured from 20 simulation realizations 
so as to mimic 
the
measurement accuracies for a survey of 2000 sq. degrees coverage and with
redshift range $0.2<z<0.4$ (see below for more details).
However, there
are two complications that in reality need to be taken into account: 
 the contribution
to $\vlos$ from the cosmological redshift, which is given by
$H \D r_{\rm los}$, where $\D r_{\rm los}$ is the line-of-sight separation
between the galaxy and the cluster; and the contribution from motions
of galaxies within their parent halos.  The Hubble flow contribution
can be modeled if the real-space cluster-galaxy correlation function
on scales of interest is known.  In practice, we can only measure 
the \emph{redshift-space} correlation function, which in turn receives
contributions from the velocities
-- this greatly complicates the subtraction of the Hubble flow contribution.  
One approach to solve this considerable difficulty is to construct
a joint model of the $\vlos-r_p$ phase space and the redshift-space
correlation function.  Another possibility is to measure stacked
weak lensing around the galaxies, which
yields the real-space galaxy-matter correlation function.
Combining the cluster-matter and galaxy-matter correlation 
can be used to infer the galaxy-cluster correlation function.  

Another effect which has to be included is 
the motion of galaxies relative to the center-of-mass of their parent
halos.  In order
to include this contribution, we need to know the distribution of
relative velocities as well as radial offsets relative to their halos.  
If the galaxies are
dynamically relaxed within the halos, these distributions are
related by the virial theorem.  Stacked weak lensing measured for the
galaxy sample yields the mean parent halo mass as well as giving
clues to the distribution of radial offsets.  This can be used
to constrain the galaxy motions within halos \cite{HikageEtal}.

One further advantage of the stacking procedure and of considering
scales of several Mpc is that we do not necessarily require a high
number density of spectroscopic galaxies.  In contrast, deep dedicated
observations would be needed if one were to determine the velocity 
dispersion of individual clusters.  
The stacked weak lensing measurement requires an adequately deep imaging survey
so that images of background galaxies are well resolved.  

\begin{figure}[t]
\centering
\includegraphics[angle=-90,width=0.44\textwidth]{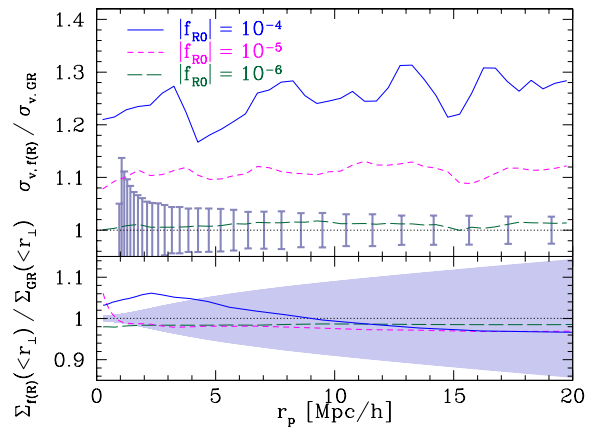}
\caption{\textit{Upper panel:} Ratio of the velocity dispersion $\s_v$
along the line of sight measured around halos with $M_{300} >
10^{14}\Msunh$ in $f(R)$ simulations to that measured around halos of
the same mass in $\Lambda$CDM simulations.  The error bars are estimated
from the simulations, as in Fig.~\ref{fig:ps}, for a spectroscopic
survey of 2000 sq. degrees.  \textit{Lower panel:} Ratio of the
enclosed projected mass profiles of the same halos in $f(R)$ and
$\Lambda$CDM simulations.  This is approximately what stacked lensing
would measure.  The shaed region 
indicates the range of statistical uncertainties for an imaging survey of the
same area (see text).
\label{fig:fR}}
\end{figure}
\begin{figure}[t]
\centering
\includegraphics[angle=-90,width=0.44\textwidth]{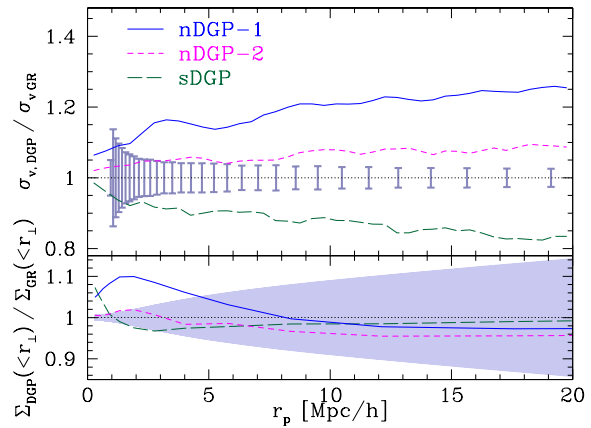}
\caption{Same as \reffig{fR}, but for DGP models (see text).
\label{fig:DGP}}
\end{figure}
{\em Results:} \label{sec:MG}
We now turn to the signatures of modified gravity in the $v_{\rm los}-r_p$
phase space.  We begin with the modified action $f(R)$ model
(see \cite{Sotiriou:2008rp} and references therein), specifically
the one of \cite{hu07a} with $n=1$.  The model can be parametrized by
the amplitude $f_{R0}$ of the scalar degree of freedom $f_R \equiv df/dR$
today, with $f_{R0}=0$ being equivalent to $\Lambda$CDM.  For the values
considered here ($|f_{R0}| = 10^{-4} - 10^{-6}$), the expansion history
is indistinguishable from $\Lambda$CDM.  Current best cosmological 
constraints are a $|f_{R0}| <$a few$\times10^{-4}$ \cite{fRcluster,LombriserfR}.  
In $f(R)$ gravity, gravitational
forces are enhanced by a factor of $4/3$ within the redshift-dependent
Compton wavelength of the field.  In addition, this model incorporates
the chameleon mechanism which restores GR in high-density environments
\cite{khoury04a,hu07a}.  
The upper panel of \reffig{fR} 
 shows the dispersion $\s_v$ of the line-of-sight velocity distribution 
in bins of $r_p$ measured around halos above $10^{14}\Msunh$ in $f(R)$ 
N-body simulations
\cite{HPMpaper,HPMhalopaper}, relative to that measured in $\Lambda$CDM
simulations around halos above the same mass threshold \footnote{Although
this measurement was done at $z=0$, a comparison with the results at $z=1$ 
shows that the enhancement in $\s_v$ is only weakly redshift-dependent.}.     
Due to the limited volume and resolution of the modified gravity simulations, we
performed this measurement for dark matter particles.  
Note that
the enhancements in $\s_v$ can become significantly larger than the effect
on the virial velocities, which are enhanced by up to a factor of 
$\sqrt{4/3} \approx 1.15$ in $f(R)$ \cite{Schmidt10}.  

In case of chameleon theories
such as $f(R)$, if the spectroscopic galaxies are screened, we expect the
enhancement of velocities to be suppressed \cite{HuiNicolisStubbs}.  
Secondary halos with $M_{300} > 3\times 10^{13}\Msunh$ identified
in the $f(R)$ simulations indicate a somewhat suppressed effect 
on the velocity dispersion for $f_{R0} \leq 10^{-5}$,
although the error bars are large.    
This suppression is consistent with the mass thresholds $\sim 10^{14}\Msunh$ and below
for the chameleon mechanism for these field values \cite{Schmidt10}.    
On the other hand, the chameleon-screening of the clusters only affects
the phase space at separations of order the virial radius of
the cluster halos, i.e. a few Mpc or less.  This can be seen for the
cases of $f_{R0}=10^{-5}, 10^{-6}$ in \reffig{fR}, for which the primary 
halos are screened in the simulations.  

\reffig{DGP} shows the same measurement in N-body simulations of the 
Dvali-Gabadadze-Porrati (DGP)
type braneworld models \cite{DGP1,Deffayet01}.  
We consider simulations for a self-accelerating DGP model
without any $\Lambda$ or dark energy ({\em sDGP}, \cite{DGPMpaper}), and normal-branch
models including a dark energy component ({\em nDGP}, \cite{DGPMpaperII}).  
The dark energy equation
of state is adjusted to yield a $\Lambda$CDM
expansion history, making these nDGP models indistinguishable by 
geometric probes \cite{DGPMpaperII}.  In case of sDGP, we compare to
a GR model with an effective dark energy yielding the same expansion history,
in order to isolate the modified structure growth effects.  
DGP models are characterized by the cross-over scale $r_c$,
above which gravity transitions from 4D to 5D.  On scales below $r_c$, 
gravity is described by a 4D scalar-tensor theory, where the strength of
the modified force scales with $H r_c$.  In sDGP, 
$r_c = 1.35 H_0^{-1} = 4038\Mpch = 6118$~Mpc,  
while in nDGP--1 (--2) it is taken to be
500 (3000) Mpc.  As expected, we see that sDGP yields smaller velocities
than GR, since gravity is weakened in the self-accelerating branch.  
Conversely, normal-branch models yield higher velocities.  We find
no indication of a suppression of the effect when considering secondary
halos ($M_s > 3\times 10^{13}\Msunh$) instead of dark matter, consistent 
with the fact that the 
Vainshtein screening mechanism inherent in these braneworld scenarios does 
not directly lead to a velocity bias \cite{HuiNicolisStubbs}.

Will upcoming surveys be able to detect such modified gravity signatures in the 
phase space distribution?  
The statistical uncertainties in $\s_v$ arise from an imperfect
sampling due to a finite number of the cluster-galaxy
pairs and from cosmic variance due to a finite volume coverage.  To
make realistic forecasts, we adopt survey parameters that resemble the
planned imaging and spectroscopic surveys with the Subaru Telescope
\cite{HikageEtal};
we assume a survey area of $2000$ square degrees, and consider as
cluster sample halos with mass greater than
$10^{14}M_\odot/h$ and in the redshift range $0.2<z<0.4$.  The comoving
volume corresponds to 0.23~$({\rm Gpc}/h)^3$.  We chose the mass range
so that the massive halos allow an accurate measurement of the average
mass profile
with weak lensing \cite{OguriTakada:11}.  We choose a cluster sample
at relatively low redshifts 
to allow for a 
denser sampling of redshifts of the secondary halos
(galaxies).  For the latter, we assume that the galaxies reside 
in halos 
with masses
$3\times 10^{13}\le M_{s}<10^{14}M_\odot/h$ as in Fig.~\ref{fig:ps}, and
assume one galaxy per halo 
residing 
at the halo's center of mass.  
The mean number densities of the
primary and secondary halos, found from the $\Lambda$CDM simulations,
are $1.7\times 10^{-5}$ and $8.3\times 10^{-5}~[{{\rm Mpc}/h}]^{-3}$,
respectively, the latter being lower than the density of
spectroscopic galaxies for the SDSS BOSS survey \cite{Whiteetal:11} or
the target density for the Subaru survey.

To estimate the measurement accuracies, we divide each of our 
20 realizations of N-body simulations for $\Lambda$CDM into 27 subvolumes 
of $0.056~({\rm Gpc}/h)^3$ at the output redshift $z=0.35$, in order to
increase the sample size.  
We compute the stacked phase space
distribution ($\vlos,r_p$) and RMS $\s_v(r_p)$ using pairs in each subvolume.  
We then compute the mean and covariance of
the $\sigma_v$-profiles for all radial bins over the 540 samples.  Finally, we
rescale the covariance by $(V_{\rm sim}/V_{\rm survey})^{1/2}$.  
The 1$\sigma$ uncertainty in each radial bin is shown in the upper 
panel of \reffig{fR} and \ref{fig:DGP}.  The constraining power of
the assumed galaxy survey is clearly very significant, over a wide range
of separations.  Note that the error bars at different radial bins are highly
correlated.  
To be more quantitative, we can estimate the value of
$-2 \Delta\ln{\cal L}$ between the $\Lambda$CDM and $f(R)$
models, using the full covariance of $\s_v(r_p)$ as measured in the 
$\Lambda$CDM simulations.  This yields 
$218,\  70$ and $2.2$ for 
$|f_{R0}|=10^{-4}, 10^{-5}$ and $10^{-6}$, respectively, 
assuming that the shape of the velocity profile
is perfectly known.  This shows that there is enough signal-to-noise to
probe $f(R)$ gravity down to field values at which the secondary halos
become chameleon-screened and this measurement loses its power.  
Adding a log-normal scatter in mass of $\sigma(\ln(M))=0.2$ in both
primary and secondary halos changes the velocity profile by
less than 5\%.  Hence the velocity profile appears to be robust with respect
to uncertainties in mass estimates of the halos.

For comparison, the lower panels of \reffig{fR} and \ref{fig:DGP} show the 
ratio of the enclosed projected mass profiles
around the primary halos in modified gravity to that in $\Lambda$CDM.  
This quantity can be reconstructed from weak lensing measurements, and
has been used to constrain $f(R)$ gravity \cite{LombriserEtal11}.  
Clearly, the departures in the mass profile are much smaller than those
in the velocities.  The range enclosed by the two thin-solid curves 
shows the expected
$1\sigma$ measurement uncertainties for a Subaru-type imaging survey
covering the same region of the sky, i.e. 2000 square degrees. The
lensing errors are determined by the survey area and the shot noise \cite{OguriTakada:11};
we assumed a background galaxy density at $z_s>0.6$ of $\bar{n}_g=22$~arcmin$^{-2}$
and a RMS intrinsic ellipticity of $\sigma_{\epsilon}=0.22$.  Given the size
of the error bars relative to the modified gravity effects, it is clear
the lensing signal itself is a much less powerful probe of gravity than velocities.  

\label{sec:concl}

{\em Discussion}: In this {\em Letter}, we have investigated a method of using
the phase space distribution around massive clusters to constrain
modified gravity models.  Using collisionless numerical simulations for
$\Lambda$CDM, $f(R)$ and DGP models, we demonstrated that the velocity 
dispersion as a function of transverse separation shows up to order unity 
deviations when compared to the profile in GR.  On the other hand, 
the effect on the interior mass profile,
which is measurable through stacked weak lensing, is much less
affected by modifications to gravity.  
While we have concentrated on the second moment of the velocity distribution
here, in principle even more information is contained in the higher moments.  
As working examples, we showed that a spectroscopic survey covering an
area of 2000 square degrees can in principle yield greatly improved 
constraints on $f(R)$ and DGP models (see Figs.~\ref{fig:fR} and \ref{fig:DGP}).  The scales probed by this
method are in the (weakly) nonlinear regime, and bridge the gap between the
scales probed by redshift-space distortions in galaxy two-point correlations
on large scales and virial velocities within
halos on small scales.  By combining these different methods, we can probe
gravity properties over a wide range of scales, and have a
better chance of capturing the signatures of the screening mechanisms, 
should the accelerated expansion in fact be
due to the breakdown of Einstein gravity on cosmological scales. \\


TYL, TN and MT are  supported in part by Grant-in-Aid for Young
Scientists (Nos. 22740149 and 23340061) and by WPI Initiative, MEXT, Japan.
Numerical calculations for the present work have been 
in part carried out under the 
“Interdisciplinary Computational Science Program” in Center 
for Computational Sciences, 
University of Tsukuba, and on Cray XT4 at Center for Computational
Astrophysics, CfCA, of National Astronomical Observatory of Japan.
FS is supported by the Gordon and Betty Moore Foundation at Caltech.  
MT is also supported
by the FIRST program
``Subaru Measurements of Images and Redshifts (SuMIRe)'', CSTP, Japan.
\bibliography{DGPM}

\end{document}